# Layout considerations for a future electron plasma research accelerator facility EuPRAXIA

P A Walker[1], R W Assmann[1], R Brinkmann[1] E Chiadroni[2], U Dorda[1], M Ferrario[2], D Kocon[3], B Marchetti[1], L Pribyl[3], A Specka[4], R Walczak[5,6]

1 Deutsches Elektronen-Synchrotron, 22607 Hamburg, Germany
2 INFN, Laboratori Nazionali di Frascati, 00044 Frascati, Rome, Italy
3 ELI-Beamlines, Dolni Brezany, Czech Republic
4 LLR, CNRS, Ecole Polytechnique, Palaiseau and Université Paris Saclay, France
5 John Adams Institute, Oxford University, UK
6 University of Oxford, Oxford OX1 2JD, UK

**Abstract**:
The Horizon 2020 Project EuPRAXIA ("European Plasma Research Accelerator with eXcellence In Applications") is preparing a conceptual design for a highly compact and cost-effective European facility with multi-GeV electron beams using plasma as the acceleration medium. The design includes two user areas: one for FEL science and one for High Energy Physics (HEP) detector development and other pilot applications. The accelerator facility will be based on a laser and/or a beam driven plasma acceleration approach. This contribution introduces layout considerations of the future plasma accelerator facilities in the context of EuPRAXIA. It compares conventional and novel plasma accelerator facility requirements and presents potential layouts for the future site. Together with performance analysis, cost effectiveness, and targeted user cases of the individual configurations, such layout studies will later enable a ranking of potential configurations. Based on this information the optimal combination of technologies will be defined for the 2019 conceptual design report of the EuPRAXIA facility.

## 1. Introduction

16 European partner laboratories and additional 22 associated partners from the EU, Israel, China, Japan, Russia and the USA [1, 2] have formed the EuPRAXIA collaboration. EuPRAXIA is structured into 14 working packages of which eight work packages ("WP") receive direct EU funding and their topics include: plasma and laser simulations (WP2), plasma accelerator structures (WP3), laser design (WP4), conventional beam physics (WP5), FEL radiation (WP6), and a table-top test beam for HEP and other applications (WP7). WP1 and WP8 concentrate on management and outreach to the public, respectively. In-kind work packages (WP9 - WP14) include additional approaches: beam driven plasma acceleration PWFA (WP9), hybrid acceleration schemes (WP14), alternative radiation generation (WP13) and alternative laser sources such as fiber lasers (WP10). WP11 and WP12 connect to prototyping on plasma-based FEL's and facility access for experiments until 2019. Partners from industry are Amplitude Technologies, Thales and Trumpf Scientific which contribute their experience towards a successful completion of the design report.

## 2. Plasma acceleration

Plasma acceleration has been first proposed by Veksler [2] and Tajima and Dawson [3] decades ago. Both electron beams (plasma wakefield acceleration, PWFA) or intense laser pulses (laser wakefield acceleration, LWFA) are well suited for accelerating charged particles [4] by creating longitudinal plasma waves. While many advances have been achieved over the last two decades [5-25], both within the PWFA and LWFA community, this paper concentrates on combining the advances of both within in one facility. It describes the current approach to plan a facility based on the best available technology and available simulations.

## 3. Layout considerations

Because both laser-driven and beam-driven approaches as well as combined plasma acceleration schemes - using LWFA-produced beams as drivers of PWFA stages [29, 30] - are considered at this point of the EuPRAXIA study, the design of the facility considers all of these techniques. The final EuPRAXIA facility design will include either one or several of the proposed options depending on the available funding and the science targeted.

The first iteration of the design parameters to provide a 5 GeV beam, FEL radiation and other pilot applications such as positron production, detector tests, and compact X-ray sources, was published in October 2016 [31]. A more detailed description of all goal parameters can be found in [1, 31]. The different configurations to achieve these goals are:

> Configuration 1: LWFA with internal injection;
> Configuration 2: LWFA with external injection from an RF accelerator;
> Configuration 3: LWFA with external injection from a laser plasma injector;
> Configuration 4: PWFA with an RF electron beam; and
> Configuration 5: PWFA with LWFA produced electron beam (hybrid schemes).

These configurations of a potential layout of the EuPRAXIA accelerator tunnel [43] are shown in Figure 1, excluding user areas. Configurations 1 to 4 are visualized with configuration 5 being able to be implemented in configuration 1 to 4. All RF and laser infrastructure is being supplied from the level above (laser paths are shown in red) and undulators are shown in the bottom right corners (yellow).

Figure 1a depicts configuration 1: LWFA with internal injection in which two plasma stages are included supplied with two laser beams (red). Figure1b shows configuration 2: LWFA with external injection from an RF accelerator. The RF gun and S-band structures are shown in front of a dogleg which transports the electrons to the two plasma stages. In addition to the S-band structures, where acceleration is achieved with S-band sections (more than the 3 shown) and with an additional chicane, X-band structures could be used behind the S-band structure to accelerate the beam up to 540 MeV before the beam is transported to the plasma cell. Configuration 3 is shown in Figure 1c: LWFA with external injection from a laser plasma injector. The externally injected electron beam could be supplied by the laser plasma injector, which is driven by a separate laser (faint red line). And Figure 1d depicts configuration 4: beam driven plasma acceleration. Using the same infrastructure of RF gun and S-band structure, the PWFA case uses additional X-band structures to accelerate beams to ~500 MeV before using it inside a single plasma accelerator stage. The footprint of the accelerator tunnel could be up to 5 times smaller than in conventional accelerator facilities.

The two-level layout of the building is shown in Figure 2, in which the RF and laser infrastructure for RF and plasma acceleration cavities on the first floor is shown. On a third level (not shown) other infrastructure such as heating, cooling and electricity supplies could be located. EuPRAXIA is a site-independent design study and potential sites that have been discussed are: EuPRAXIA@SPARC_LAB (Frascati, Italy), SINBAD (Hamburg, Germany), CILEX (Paris, France), CLF (Didcot, UK) and ELI_beams (Prague, Czech Republic).

## 4. Summary

The EuPRAXIA collaboration is preparing a conceptual design report for a multi-GeV plasma-based accelerator with outstanding beam quality. The facility design aims to include: FEL radiation in the soft (to hard) X-ray range, a table-top test beam for HEP detectors and industry, and a compact X-ray source for medical imaging. Both laser and electron beams are considered as power sources for the plasma accelerator and a potential layout is presented. The final facility layout may only include one or two configurations dependent on user needs and funding available. This decision will be made in 2019.

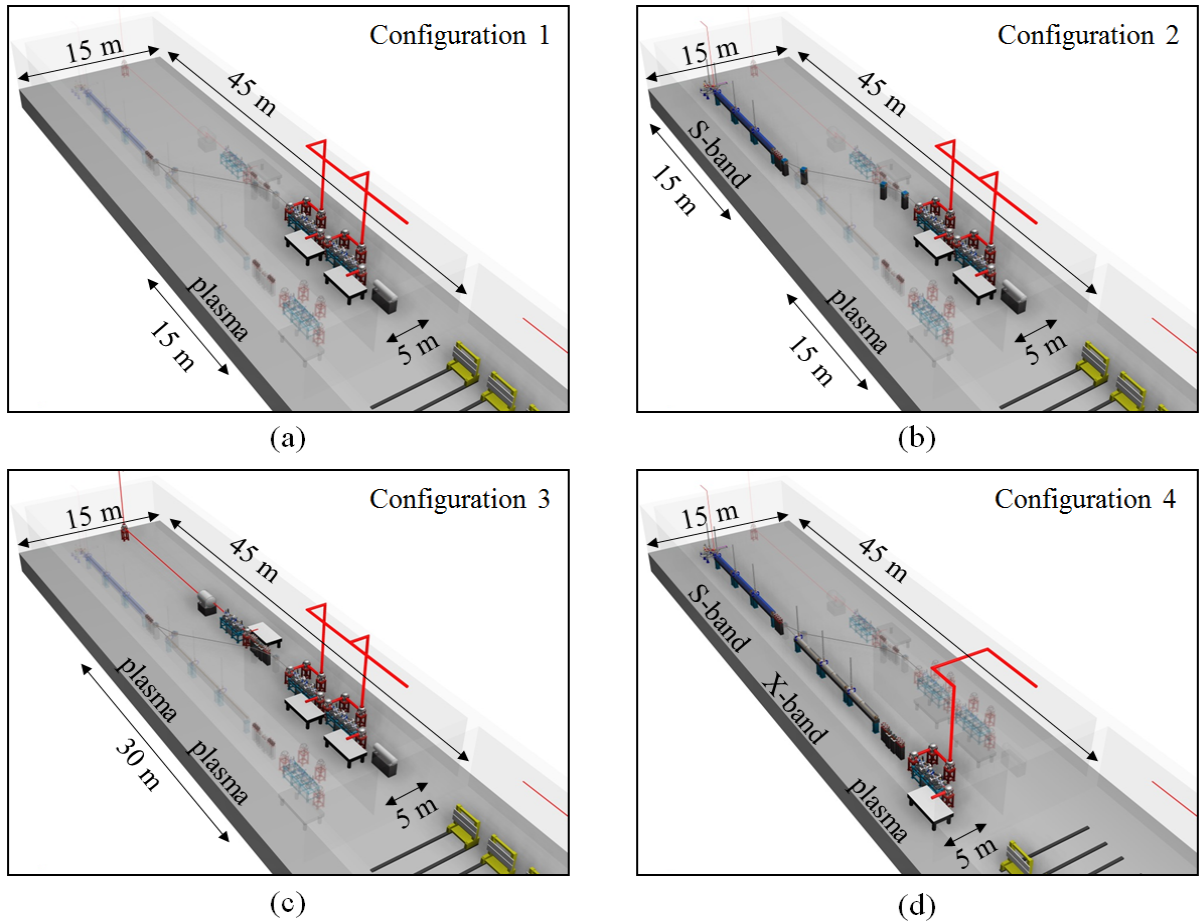

**Figure 1.** The preliminary layout of the EUPRAXIA accelerator tunnel is shown. Four of the five configurations are depictured, with the configuration 5 being able to be implemented in all previous configurations.

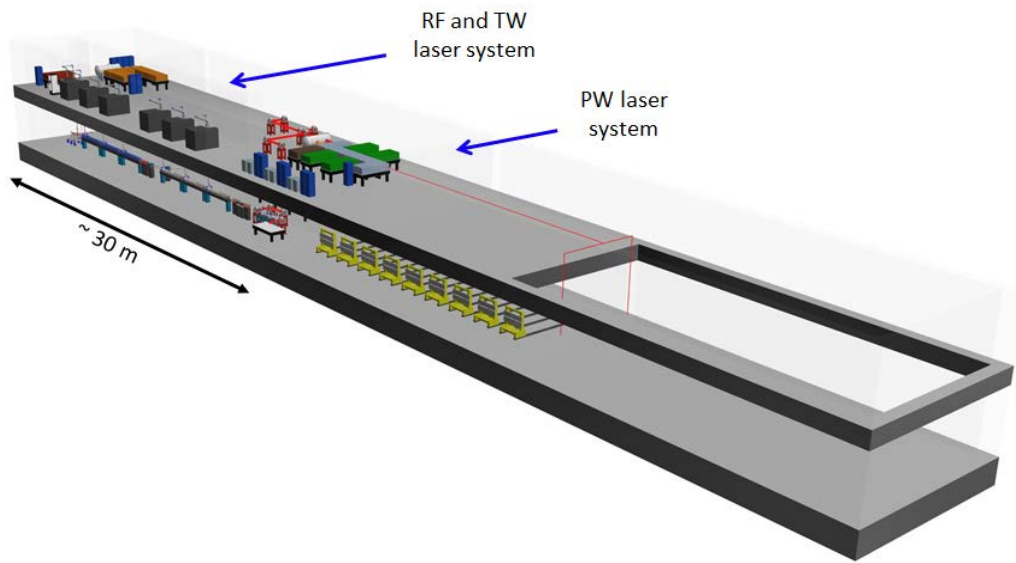

**Figure 2.** The two level structure for the preliminary EuPRAXIA layout is shown: above the accelerator tunnel, with RF, plasma cavities, and undulators, the infrastructure of the RF and laser power sources us positioned. The high power laser will be available in the user areas, as indicated by the faint red line.

**Acknowledgments**

This work was supported by the European Union's Horizon 2020 research and innovation programme under grant agreement No. 653782.